\documentclass[onecolumn,showpacs,preprintnumbers,amsmath,amssymb]{revtex4-1}
\usepackage{amsmath,amsfonts,amssymb}
\usepackage{graphicx}

\begin{document}
\title{Black hole solutions in de Rham-Gabadadze-Tolley massive gravity}
\author{Ping Li}
\author{Xin-zhou Li}
\email{kychz@shnu.edu.cn}
\author{Ping Xi}
\email{xiping@shnu.edu.cn}
\affiliation{Center for Astrophysics, Shanghai Normal University, 100 Guilin Road, Shanghai, 200234, China}
\date{\today}

\begin{abstract}
We present a detailed study of the static spherically symmetric solutions in de Rham-Gabadadze-Tolley (dRGT) theory. Since the diffeomorphism invariance can be restored by introducing the St\"{u}ckelberg fields $\phi^a$, there is new invariant $I^{ab}=g^{\mu\nu}\partial_{\mu}\phi^a\partial_\nu\phi^b$ in the massive gravity, which adds to the ones usually encountered in general relativity (GR). In the unitary gauge $\phi^a=x^\mu\delta_\mu^a$, any inverse metric $g^{\mu\nu}$ that has divergence including the coordinate singularity in GR would exhibit a singularity in the invariant $I^{ab}$. Therefore, there is no conventional Schwarzschild metric if we choose unitary gauge. In this paper, we obtain a self-consistent static spherically symmetric ansatz in the nonunitary gauge. Under this ansatz, we find that there are seven solutions including the Schwarzschild solution, Reissner-Nordstr\"{o}m solution and five other solutions. These solutions may possess an event horizon depending upon the physical parameters (Schwarzschild radius $r_s$, scalar charge $S$ and/or electric charge $Q$). If these solutions possess an event horizon, we show that the singularity of $I^{ab}$ is absent at the horizon. Therefore, these solutions may become candidates for black holes in dRGT.
\end{abstract}
\pacs{04.50.Kd, 14.70.Kv}
\maketitle

\section{Introduction}
It is an interesting question whether general relativity (GR) is a solitary theory from both the theoretical and phenomenological sides. One of the modifying gravity theories is the massive deformation of GR. A comprehensive review of massive gravity can be found in \cite{Rham}. We can divide the massive gravity theories into two varieties: the Lorentz invariant type (LI) and the Lorentz breaking type (LB). Though for many years it was certain that the theory of LI massive gravity always contains the Boulware-Deser (BD) ghosts \cite{Boulware}, a kind of its nonlinear extension was recently constructed by de Rham, Gabadadze, and Tolley (dRGT) \cite{Gabadadze,Rham1,Rham2,Rham3}. In GR, the spherically symmetric vacuum solution to the Einstein equation is a benchmark, and its massive deformation also plays a crucial role in massive gravity. A detailed study of the spherically symmetric solutions is presented in LB massive gravity \cite{Li}. dRGT theory also contains a vacuum solution that recovers exactly a Schwarzschild-de Sitter solution \cite{Nieuwenhuizen,Koyama,Berezhiani} in a group of specially selected parameters. They concentrate on a special family of dRGT theory in which there is the following relation between the two free parameters: $9\alpha_3^2+3\alpha_3-12\alpha_4+1=0$. This choice was first shown by Nieuwenhuizen \cite{Nieuwenhuizen} (see also \cite{Gruzinov}). According to the method of Koyama, Niz and Tasinato \cite{Koyama}, one can start in the unitary gauge $\phi^a=x^\mu\delta_\mu^a$, and consider a most general stationary spherically symmetric metric. The resulting metric is the Schwarzschild-de Sitter solution in the Gullstrand-Painlev\'{e} coordinates \cite{Berezhiani}. This metric can be transformed to a static slicing by a suitable coordinate transformation, in which the metric is accompanied by nontrivial backgrounds for the St\"{u}ckelberg fields \cite{Berezhiani}. Other analyses and phenomenological studies were discussed in Refs. \cite{Koyama1,Sbisa,Comelli,Chkareuli,Sjors,Cai} for the black holes and spherically symmetric solutions in dRGT.

The symmetric tensor field $h_{\mu\nu}\equiv g_{\mu\nu}-\eta_{\mu\nu}$ is the gravitational analogue to the Proca field in the massive electrodynamics, describing all five modes of the massive graviton. The diffeomorphism invariance can be restored by introducing the four St\"{u}ckelberg fields \cite{Hinterbichler} and replacing the Minkowski metric by the covariant tensor $\partial_\mu\phi^a\partial_\nu\phi^b\eta_{ab}$, then the symmetric tensor $H_{\mu\nu}$ describes the covariantized metric perturbation. In the unitary gauge, $H_{\mu\nu}$ reduces to $h_{\mu\nu}$. There is a new basic invariant $I^{ab}=g^{\mu\nu}\partial_\mu\phi^a\partial_\nu\phi^b$ in the massive gravity in addition to the ones usually encountered in GR since the existence of the four scalar fields $\phi^a$. In the unitary gauge, we have $I^{ab}=g^{\mu\nu}\delta_\mu^a\delta_\nu^b$. It is obvious that $I^{ab}$ will exhibit a singularity if $g^{\mu\nu}$ has any divergence including the coordinate singularity for the unitary gauge. De Rham and his colleagues \cite{Berezhiani} have pointed out that one would expect the singularities in $I^{ab}$ to be a problem for fluctuations around classical solutions exhibiting it. For this reason, they propose that the solution comes true only if $I^{ab}$ is nonsingular. In this paper, we continue to use this conservative rule.

As a corollary of the above point of view, there is no conventional Schwarzschild metric of massive gravity in unitary gauge, which gives rise to the following paradox. According to the vainshtein mechanism \cite{Vainshtein}, this solution of massive gravity should approximate one of GR better and better when we increase the mass of the source. That is to say, this black hole of massive gravity near its horizon should be very similar to that of GR. However, this metric would be singular at the horizon according to the argument above. Whether or not there is the conventional Schwarzschild solution in dRGT with two free parameters is one of the questions that motivates this paper.

In this work, we discuss the black holes and static spherically symmetric solutions in dRGT, where two parameters are freely chosen. Furthermore, we release from the limitations of the unitary gauge $\phi^a=x^\mu\delta_\mu^a$, and the St\"{u}ckelberg field $\phi^i$ is taken as a "hedgehog" configuration $\phi^i=\phi(r)\frac{x^i}{r}$ and $\phi^0=t+h(r)$ \cite{Li}. We have found a wide class of static spherically symmetric solutions including the Schwarzschild solution, the Reissner-Nordstr\"{o}m solution, the furry black hole solution and some new solutions. On the obtained solutions the singularities in the invariant $I^{ab}$ are absent except for the physical singularity $r=0$, so that these solutions may be regarded as candidates for modified black holes in dRGT.

The paper is organized as follows: Sec. II gives a brief review of dRGT theory \cite{Rham3}. In Sec. III, we present a self-consistent static spherically symmetric ansatz with a nonunitary gauge. In Sec. IV, we find the Schwarzschild solution and two other solutions depending upon the parameters $r_s$ and $S$, and in Sec. V the Reissner-Nordstr\"{o}m solution and three other solutions depending upon the parameters $r_s$, $Q$ and/or $S$. Moreover, the obtained solutions have nonsingular $I^{ab}$. In Appendix A, we give the expression of $I^{ab}$ under the self-consistent ansatz, and show the obtained solutions in which the singularities in the invariant $I^{ab}$ are absent except the physical singularity $r=0$.

\section{The modified Einstein equations in dRGT theory}
The gravitational action is
\begin{equation}\label{1}
S=\frac{M_{pl}^2}{2}\int d^4x\sqrt{-g}[R+m^2U(g,\phi^a)],
\end{equation}
where $R$ is the Ricci scalar, and $U$ is a potential for the graviton which modifies the gravitational sector. The potential is composed of three parts,
\begin{equation}\label{2}
U(g,\phi^a)=U_2+\alpha_3U_3+\alpha_4U_4,
\end{equation}
where $\alpha_3$ and $\alpha_4$ are dimensionless parameters, and
\begin{eqnarray}\label{3}
U_2&=&[\mathcal{K}]^2-[\mathcal{K}^2],\nonumber\\
U_3&=&[\mathcal{K}]^3-3[\mathcal{K}][\mathcal{K}^2]+2[\mathcal{K}^3],\\
U_4&=&[\mathcal{K}]^4-6[\mathcal{K}]^2[\mathcal{K}^2]+8[\mathcal{K}][\mathcal{K}^3]+3[\mathcal{K}^2]^2-6[\mathcal{K}^4].\nonumber
\end{eqnarray}
Here the square brackets denote the traces, i. e., $[\mathcal{K}]=\mathcal{K}^\mu_{\ \mu}$ and
\begin{eqnarray}\label{4}
\mathcal{K}^\mu_{\ \nu}&=&\delta^\mu_{\ \nu}-\sqrt{g^{\mu\alpha}\partial_\alpha\phi^a\partial_\nu\phi^b\eta_{ab}}\nonumber\\
&\equiv&\delta^\mu_{\ \nu}-\sqrt{\mathbf{\Sigma}}^\mu_{\ \nu}
\end{eqnarray}
where the matrix square root is $\sqrt{\mathbf{\Sigma}}^\mu_{\ \alpha}\sqrt{\mathbf{\Sigma}}^\alpha_{\ \nu}=\mathbf{\Sigma}^\mu_{\ \nu}$, $g^{\mu\nu}$ is the physical metric, $\eta_{ab}$ is the reference metric and $\phi^a$ are the St\"{u}ckelberg scalars introduced to restore general covariance \cite{Arkani}.

Variation of the action with respect to the metric leads to the modified Einstein equations
\begin{equation}\label{5}
G_{\mu\nu}+m^2T^{(\mathcal{K})}_{\mu\nu}=\frac{1}{M_{pl}^2}T^{(m)}_{\mu\nu},
\end{equation}
where
\begin{equation}\label{6}
T^{(\mathcal{K})}_{\mu\nu}=\sqrt{-g}\frac{\delta(\sqrt{-g}U)}{\delta g^{\mu\nu}}.
\end{equation}
From (\ref{4}), we have
\begin{equation}\label{7}
\mathcal{K}^{n\ \mu}_{\ \ \ \ \nu}=\delta^\mu_{\ \nu}+\Sigma_{k=1}^{n}(-1)^k(^n_k)\mathbf{\Sigma}^{\frac{k}{2}\ \mu}_{\ \ \ \ \nu}.
\end{equation}
Thus, $[\mathcal{K}^n]$ can be written as follows,
\begin{eqnarray}\label{8}
[\mathcal{K}]=4-[\mathbf{\Sigma}],\nonumber
\end{eqnarray}
\begin{eqnarray}
[\mathcal{K}^2]=4-2[\sqrt{\mathbf{\Sigma}}]+[\mathbf{\Sigma}],\nonumber
\end{eqnarray}
\begin{eqnarray}
[\mathcal{K}^3]&=&4-3[\sqrt{\mathbf{\Sigma}}]+3[\mathbf{\Sigma}]-[\mathbf{\Sigma}^{\frac{3}{2}}],
\end{eqnarray}
\begin{eqnarray}
[\mathcal{K}^4]&=&4-4[\sqrt{\mathbf{\Sigma}}]+6[\mathbf{\Sigma}]-4[\mathbf{\Sigma}^{\frac{3}{2}}]+[\mathbf{\Sigma}^2].\nonumber
\end{eqnarray}

The symmetric tensor $H_{\mu\nu}$ describes the covariantized metric perturbation, which reduces to the $h_{\mu\nu}$ in the unitary gauge. Therefore, it is natural to split $\phi^a$ into two parts: $\phi^a=x^a-\pi^a$ and $\pi^a=0$ in the unitary gauge. It is useful that we adopt the following decomposition in the nonunitary gauge,
\begin{equation}\label{9}
\pi^a=\frac{mA^a+\partial^a\pi}{\Lambda^3},
\end{equation}
where $A^\mu$ describes the helicity $\pm1$, $\pi$ is the longitudinal mode of the graviton in the decoupling limit \cite{Berezhiani}. Moreover, $M_{pl}\rightarrow\infty$ and $m\rightarrow0$ in the decoupling limit \cite{Arkani}, while $\Lambda^3\equiv M_{pl}m^2$ is held fixed. This limit represents the approximation in which the energy scale $E$ is much greater than the graviton mass scale.

\section{A self-consistent spherically symmetric ansatz}
The invariant $I^{ab}=g^{\mu\nu}\partial_\mu\phi^a\partial_\nu\phi^b$ will exhibit a singularity if $g^{\mu\nu}$ has any divergence including the coordinate singularity in the unitary gauge. Therefore, we consider the static spherically symmetric ansatz as follows,
\begin{eqnarray}\label{10}
&ds^2&=-b^2(r)dt^2+a^2(r)dr^2+r^2d\Omega^2,\nonumber\\
&\phi^0&=t+h(r),\\
&\phi^i&=\phi(r)\frac{x^i}{r},\nonumber
\end{eqnarray}
where $\phi^i$ is taken as a "hedgehog" configuration. The St\"{u}ckelberg fields reduce to the unitary gauge $\phi^a=x^\mu\delta^a_\mu$ only if $h(r)=0$ and $\phi(r)=r$ in the ansatz (\ref{10}). The ansatz (\ref{10}) contains two additional radial functions, $h(r)$ and $\phi(r)$ as compared with GR. In GR, Birkhoff proved that the vacuum solution of the Einstein field equations in the region exterior to the source is still stationary and is still the Schwarzschild solution \cite{Birkhoff}. The result is known as Birkhoff's theorem. The standard Birkhoff's theorem is absent, in general, for massive gravity since the $T^{(\mathcal{K})}_{\mu\nu}$ term has modified the Einstein equations. Therefore, the self-consistency of (\ref{10}) imposes restrictions on $h(r)$ and $\phi(r)$, that is, $h(r)$ and $\phi(r)$ should be solutions of the modified Einstein equations (\ref{5}).

Under the ansatz (\ref{10}), the matrix $\mathbf{\Sigma}=(\mathbf{\Sigma}^{\mu}_{\ \nu})$ takes the form
\begin{eqnarray}\label{11}
\mathbf{\Sigma}=
\left(\begin{array}{cccc}
\frac{1}{b^2}&\frac{h'}{b^2}&0&0\\
-\frac{h'}{a^2}&\frac{\phi'^2-h'^2}{a^2}&0&0\\
0&0&\frac{\phi^2}{r^2}&0\\
0&0&0&\frac{\phi^2}{r^2}
\end{array}\right),
\end{eqnarray}
where primes denote derivatives with respect to $r$. For a $2\times2$ matrix $\mathbf{M}$, the Cayley-Hamilton theorem tell us that
\begin{equation}\label{12}
[\mathbf{M}]\mathbf{M}=\mathbf{M}^2+(\det{\mathbf{M}})\mathbf{I}_2,
\end{equation}
where $\mathbf{I}_2$ is $2\times2$ identity matrix. We define $\mathbf{\Sigma}_2$ as the upper left-hand $2\times2$ submatrix of $\mathbf{\Sigma}$, and use $\det{\mathbf{M}^n}=(\det{\mathbf{M}})^n$ to find the square root of $\mathbf{\Sigma}_2$ ,
\begin{eqnarray}\label{13}
\sqrt{\mathbf{\Sigma}_2}=\frac{1}{[\sqrt{\mathbf{\Sigma}_2}]}
\left(\begin{array}{cc}
\frac{1}{b^2}+\frac{\alpha\phi'}{ab}&\frac{h'}{b^2}\\
-\frac{h'}{a^2}&\frac{\phi'^2-h'^2}{a^2}+\frac{\alpha\phi'}{ab}
\end{array}\right),
\end{eqnarray}
where $\alpha=\mathrm{sgn}(\frac{\phi'}{ab})$ and
\begin{equation}\label{14}
[\sqrt{\mathbf{\Sigma}_2}]=\sqrt{(\frac{\phi'}{a}+\frac{\alpha}{b})^2-(\frac{h'}{a})^2}.
\end{equation}
Using (\ref{11}), (\ref{13}) and (\ref{14}), we have
\begin{eqnarray}\label{15}
\sqrt{\mathbf{\Sigma}}=
\left(\begin{array}{cc}
\sqrt{\mathbf{\Sigma}_2}&0\\
0&\frac{\phi}{r}\mathbf{I}_2
\end{array}\right),
\end{eqnarray}
\begin{eqnarray}\label{16}
\mathbf{\Sigma}=
\left(\begin{array}{cc}
\mathbf{\Sigma}_2&0\\
0&\frac{\phi^2}{r^2}\mathbf{I}_2
\end{array}\right),
\end{eqnarray}
\begin{eqnarray}\label{17}
\mathbf{\Sigma}^{\frac{3}{2}}=
\left(\begin{array}{cc}
\mathbf{\Sigma}_2^{\frac{3}{2}}&0\\
0&\frac{\phi^3}{r^3}\mathbf{I}_2
\end{array}\right),
\end{eqnarray}
\begin{eqnarray}\label{18}
\mathbf{\Sigma}^2=
\left(\begin{array}{cc}
\mathbf{\Sigma}_2^2&0\\
0&\frac{\phi^4}{r^4}\mathbf{I}_2
\end{array}\right),
\end{eqnarray}
and
\begin{equation}\label{19}
[\sqrt{\mathbf{\Sigma}}]=[\sqrt{\mathbf{\Sigma}_2}]+\frac{2\phi}{r},
\end{equation}
\begin{equation}\label{20}
[\mathbf{\Sigma}]=[\sqrt{\mathbf{\Sigma}_2}]^2-\frac{2\alpha\phi'}{ab}+\frac{2\phi^2}{r^2},
\end{equation}
\begin{equation}\label{21}
[\mathbf{\Sigma}^{\frac{3}{2}}]=[\sqrt{\mathbf{\Sigma}_2}]^3-\frac{3\alpha\phi'}{ab}[\sqrt{\mathbf{\Sigma}_2}]+\frac{2\phi^3}{r^3},
\end{equation}
\begin{equation}\label{22}
[\mathbf{\Sigma}^2]=[\sqrt{\mathbf{\Sigma}_2}]^4-\frac{4\alpha\phi'}{ab}[\sqrt{\mathbf{\Sigma}_2}]^2+\frac{2\phi'^2}{a^2b^2}+\frac{2\phi^4}{r^4}.
\end{equation}
Substituting (\ref{8}) and (\ref{19})-(\ref{22}) into (\ref{3}), we obtain
\begin{equation}\label{23}
U_2=(\frac{4\phi}{r}-6)[\sqrt{\mathbf{\Sigma}_2}]+12(1-\frac{\phi}{r})+\frac{2\alpha\phi'}{ab}+\frac{2\phi^2}{r^2},
\end{equation}
\begin{equation}\label{24}
U_3=(\frac{24\phi}{r}-\frac{6\phi^2}{r^2}-18)[\sqrt{\mathbf{\Sigma}_2}]+12(2-\frac{3\phi}{r})+\frac{12\alpha\phi'}{ab}(1-\frac{\phi}{r})+\frac{12\phi^2}{r^2},
\end{equation}
\begin{equation}\label{25}
U_4=-24(1-\frac{\phi}{r})^2[\sqrt{\mathbf{\Sigma}_2}]+24(1-\frac{\phi}{r})^2(1+\frac{\alpha\phi'}{ab}).
\end{equation}
Thus, we obtain the nonzero components of $T^{(\mathcal{K})\ \mu}_{\ \ \ \ \ \ \ \nu}$ as follows
\begin{eqnarray}
T^{(\mathcal{K})\ 0}_{\ \ \ \ \ \ \ 0}&=&(1-\frac{2\phi}{r})[\sqrt{\mathbf{\Sigma}_2}]-\frac{4}{[\sqrt{\mathbf{\Sigma}_2}]}(1-\frac{\phi}{r})(\frac{1}{b^2}+\frac{\alpha\phi'}{ab})+\frac{2\phi}{r}
+\frac{\alpha\phi'}{ab}-\frac{\phi^2}{r^2}+3\alpha_3([\sqrt{\mathbf{\Sigma}_2}](\frac{\phi^2}{r^2}-1)\nonumber\\
&-&\frac{2}{[\sqrt{\mathbf{\Sigma}_2}]}(1-\frac{\phi}{r})^2(\frac{1}{b^2}+\frac{\alpha\phi'}{ab})+2(1-\frac{\phi}{r})(1+\frac{\alpha\phi'}{ab}))
+12\alpha_4((\frac{\phi}{r}-1)^2(1+\frac{\alpha\phi'}{ab}-[\sqrt{\mathbf{\Sigma}_2}])),\label{26}\\
T^{(\mathcal{K})\ 1}_{\ \ \ \ \ \ \ 1}&=&(1-\frac{2\phi}{r})[\sqrt{\mathbf{\Sigma}_2}]-\frac{4}{[\sqrt{\mathbf{\Sigma}_2}]}(1-\frac{\phi}{r})(\frac{\phi'^2-h'^2}{a^2}+\frac{\alpha\phi'}{ab})
+\frac{2\phi}{r}+\frac{\alpha\phi'}{ab}-\frac{\phi^2}{r^2}-3\alpha_3((1-\frac{\phi^2}{r^2})[\sqrt{\mathbf{\Sigma}_2}]\nonumber\\
&+&\frac{2}{[\sqrt{\mathbf{\Sigma}_2}]}(1-\frac{\phi}{r})^2(\frac{\phi'^2-h'^2}{a^2}+\frac{\alpha\phi'}{ab})-2(1-\frac{\phi}{r})(1+\frac{\alpha\phi'}{ab}))
+12\alpha_4(1-\frac{\phi}{r})^2(1+\frac{\alpha\phi'}{ab}-[\sqrt{\mathbf{\Sigma}_2}]),\label{27}\\
T^{(\mathcal{K})\ 2}_{\ \ \ \ \ \ \ 2}&=&T^{(\mathcal{K})\ 3}_{\ \ \ \ \ \ \ 3}
=[\sqrt{\mathbf{\Sigma}_2}]-\frac{2\phi}{r}-\frac{\alpha\phi'}{ab}+\frac{\phi^2}{r^2}
+3\alpha_3((1-\frac{\phi}{r})^2(2-[\sqrt{\mathbf{\Sigma}_2}]))
+12\alpha_4((1-\frac{\phi}{r})^2(1+\frac{\alpha\phi'}{ab}-[\sqrt{\mathbf{\Sigma}_2}])),\label{28}\\
T^{(\mathcal{K})\ 0}_{\ \ \ \ \ \ \ 1}&=&\frac{h'}{[\sqrt{\mathbf{\Sigma}_2}]b^2}(4(\frac{\phi}{r}-1)-6\alpha_3(\frac{\phi}{r}-1)^2).\label{29}
\end{eqnarray}

It is known to all that the component of the Einstein tensor $G^0_{\ 1}=0$ for the static spherically symmetric metric. From the modified Einstein equations (\ref{5}) and $G^0_{\ 1}=0$ in the empty space case ($T^{(m)\ \mu}_{\quad\quad\ \nu}=0$), we require that $T^{(\mathcal{K})\ 0}_{\quad\quad\ 1}$ must vanish which is a self-consistent requisition for the ansatz (\ref{10}). Therefore, we obtain the $\phi$ equation as follows
\begin{equation}\label{30}
h'(1-\frac{\phi}{r})(2+3\alpha_3(1-\frac{\phi}{r}))=0,
\end{equation}
which implies the solutions are
\begin{equation}\label{31}
\phi=\beta r,
\end{equation}
where $\beta=1$ or ($\frac{2}{3\alpha_3}+1$). Here, we have abandoned the solution $h'=0$, because there are coordinate singularities in the invariant $I^{ab}$ except the Minkowskian case (see Appendix A). Just because $U_4$ contributes to $T^{(\mathcal{K})\ 0}_{\quad\quad\ 0}$ and $T^{(\mathcal{K})\ 1}_{\quad\quad\ 1}$ on the same modality, $\beta$ is independent of the parameter $\alpha_4$. Furthermore, it is easy to prove that $T^{(\mathcal{K})\ 0}_{\quad\quad\ 0}-T^{(\mathcal{K})\ 1}_{\quad\quad1}$ is proportional to $T^{(\mathcal{K})\ 1}_{\quad\quad\ 0}$ and that $T^{(\mathcal{K})\ 0}_{\quad\quad\ 0}=T^{(\mathcal{K})\ 1}_{\quad\quad\ 1}$ implies $G^{0}_{\ 0}=G^{1}_{\ 1}$. Thus, we have $a(r)b(r)=1$.

Finally, the self-consistent ansatz should be written as
\begin{eqnarray}\label{32}
&ds^2&=-b^2(r)dt^2+a^2(r)dr^2+r^2d\Omega^2,\nonumber\\
&\phi^0&=t+h(r),\\
&\phi^i&=\beta x^i,\nonumber
\end{eqnarray}
where $a(r)=(b(r))^{-1}$. Equation (\ref{32}) makes known that we take the static spherically symmetric metric at the expense of St\"{u}ckelberg fields to away from the unitary gauge. On the other hand, the nonunitary gauge provides an opportunity to avoid the singularity in the invariant $I^{ab}$ from the divergence of $g^{\mu\nu}$ in the unitary gauge. Thus, we achieve two things at one stroke. For example, de Rham and his colleagues \cite{Rham3} have found the Schwarzschild-de Sitter solution in the $9\alpha_3^2+3\alpha_3-12\alpha_4+1=0$ case, where the St\"{u}ckelberg fields stray from the unitary gauge in the static coordinate system. Meanwhile, the singularities in the invariant $I^{ab}$ are absent. Under the self-consistent ansatz (\ref{32}), the invariant $I^{ab}$ is explicitly expressed as
\begin{eqnarray}\label{33}
&I^{00}&=b^2h'^2-\frac{1}{b^2},\nonumber\\
&I^{0i}&=I^{i0}=b^2\beta h'n^i,\\
&I^{ij}&=\beta^2\delta^{ij}+\beta^2(b^2-1)n^in^j,\nonumber
\end{eqnarray}
where $(n^1,n^2,n^3)=(\sin{\theta}\cos{\varphi},\sin{\theta}\sin{\varphi},\cos{\theta})$. In Appendix A, we explore the invariant $I^{ab}$ in some detail. Obviously, $I^{00}$ is singular if $h'=0$ and $b=0$, so that we focus our attention on the case of $\phi=\beta r$.

\section{Exact solutions}
In this section, we present a detailed study of the static spherically symmetric solutions under the ansatz (\ref{32}) in the dRGT with two free parameters. The received solutions are free of singularities except from the conventional one appearing in GR (for instance, $r=0$ in the Schwarzschild metric).

The equations of motion in empty space are written as follows:
\begin{equation}\label{34}
G^{\mu}_{\ \nu}+m^2T^{(\mathcal{K})\ \mu}_{\quad\quad\ \nu}=0,
\end{equation}
where the nonzero components of the Einstein tensor are
\begin{eqnarray}\label{35}
G^0_{\ 0}&=&-\frac{2a'}{a^3r}+\frac{1}{a^2r^2}-\frac{1}{r^2},\nonumber\\
G^1_{\ 1}&=&\frac{2b'}{a^2br}+\frac{1}{a^2r^2}-\frac{1}{r^2},\\
G^2_{\ 2}&=&G^3_{\ 3}=-\frac{a'}{a^3r}+\frac{b'}{a^2br}+\frac{b''}{a^2b}-\frac{a'b'}{a^3b}.\nonumber
\end{eqnarray}
For the $ab=1$ case, (\ref{35}) reduces to
\begin{eqnarray}\label{36}
G^0_{\ 0}&=&G^1_{\ 1}=\frac{(b^2)'}{r}+\frac{b^2-1}{r^2},\nonumber\\
G^2_{\ 2}&=&G^3_{\ 3}=\frac{(b^2)''}{2}+\frac{(b^2)'}{r}.
\end{eqnarray}

\subsection{The Schwarzschild solution in dRGT}
In the $\beta=1$ case, $[\sqrt{\mathbf{\Sigma}_2}]$ can be written as
\begin{equation}\label{37}
[\sqrt{\mathbf{\Sigma}_2}]=\sqrt{(b+\frac{1}{b})^2-(bh')^2},
\end{equation}
and the equations of motion become
\begin{eqnarray}\label{38}
\frac{(b^2)'}{r}&+&\frac{b^2-1}{r^2}+m^2(-[\sqrt{\mathbf{\Sigma}_2}]+2)=0,\\
\frac{(b^2)''}{2}&+&\frac{(b^2)'}{r}+m^2([\sqrt{\mathbf{\Sigma}_2}]-2)=0.
\end{eqnarray}
The resulting expression for the static spherically symmetric solution reads as follows:
\begin{eqnarray}\label{40}
&b^2&=\frac{1}{a^2}=1-\frac{r_s}{r},\nonumber\\
&\phi^0&=t\pm r_s\ln{(r-r_s)},\\
&\phi^i&=x^i,\nonumber
\end{eqnarray}
where $r_s$ is a integral constant. This is nothing but the Schwarzschild solution of GR for the metric. However, this metric should be accompanied by nontrivial backgrounds for the St\"{u}ckelberg fields. If $r_s=0$, (\ref{40}) goes back to the Minkowski metric and the St\"{u}ckelberg fields in the unitary gauge.

Using (\ref{33}) and (\ref{40}), we have
\begin{eqnarray}\label{41}
&I^{00}&=-(1+\frac{r_s}{r}),\nonumber\\
&I^{0i}&=I^{i0}=\pm\frac{r_s}{r}n^i,\\
&I^{ij}&=\delta^{ij}-\frac{r_s}{r}n^in^j.\nonumber
\end{eqnarray}
The singularities in the invariant $I^{ab}$ are absent except the physical singularity $r=0$, so that the Schwarzschild solution of massive gravity may be regarded as a candidate for the black hole in dRGT.

\subsection{The furry black hole solutions}
In the $\beta=\frac{2}{3\alpha_3}+1$ case, $[\sqrt{\mathbf{\Sigma_2}}]$ can be written as
\begin{equation}\label{42}
[\sqrt{\mathbf{\Sigma}_2}]=\sqrt{(\beta b+\frac{1}{b})^2-(bh')^2}.
\end{equation}
From (\ref{26})-(\ref{28}) and (\ref{36}), (\ref{34}) can be rewritten as
\begin{eqnarray}
\frac{(b^2)'}{r}&+&\frac{b^2-1}{r^2}+m^2(\lambda_1[\sqrt{\mathbf{\Sigma}_2}]+\lambda_2)=0,\label{43}\\
\frac{(b^2)''}{2}&+&\frac{(b^2)'}{r}+m^2(\lambda_3[\sqrt{\mathbf{\Sigma}_2}]+\lambda_4)=0,\label{44}
\end{eqnarray}
where
\begin{eqnarray}\label{45}
\lambda_1&=&3-\frac{16\alpha_4}{3\alpha_3^2},\nonumber\\
\lambda_2&=&-6-\frac{2}{\alpha_3}-\frac{4-96\alpha_4}{\alpha_3^2}+\frac{32\alpha_4}{9\alpha_3^3},\nonumber\\
\lambda_3&=&1-\frac{4}{3\alpha_3}-\frac{16\alpha_4}{3\alpha_3^2},\\
\lambda_4&=&-2+\frac{2}{\alpha_3}+\frac{4+96\alpha_4}{9\alpha_3^2}+\frac{32\alpha_4}{9\alpha_3^3}.\nonumber
\end{eqnarray}

There are two subcases for the $\beta=\frac{2}{3\alpha_3}+1$ case: (i) $\alpha_4\neq\frac{5}{16}\alpha_3^2-\frac{\alpha_3}{6}$ and (ii) $\alpha_4=\frac{5}{16}\alpha_3^2-\frac{\alpha_3}{6}$, which correspond to distinct types of solution. In the subcase (i), we have the solution for this system as follows
\begin{eqnarray}\label{46}
b^2&=&\frac{1}{a^2}=1-\frac{r_s}{r}-\frac{S}{r^{\lambda}}+\frac{4m^2r^2}{27\alpha_3^2},\nonumber\\
\phi^0&=&t\pm\int{[(a^2+1)^2-a^2(\frac{3\alpha_3^2(1-\lambda)S}{m^2(9\alpha_3^2-16\alpha_4)r^{\lambda+2}}+2(1+\frac{1}{3\alpha_3}))^2]^{\frac{1}{2}}dr},\nonumber\\
\phi^i&=&(\frac{2}{3\alpha_3}+1)x^i,
\end{eqnarray}
where $r_s$, $S$ are integral constants and
\begin{equation}\label{47}
\lambda=2(\frac{6\alpha_3^2+4\alpha_3}{9\alpha_3^2-16\alpha_4}-1).
\end{equation}
The metric (\ref{46}) differ from the Schwarzschild-de Sitter solution by an additional powerlike term $r^{-\lambda}$.

In the subcase (ii), we have the solution for this system as follows
\begin{eqnarray}\label{48}
b^2&=&\frac{1}{a^2}=1-\frac{r_s}{r}-\frac{S\ln{r}}{r}+\frac{4m^2r^2}{27\alpha_3^2},\nonumber\\
\phi^0&=&t\pm\int{[(a^2+1)^2-a^2(\frac{3\alpha_3^2S}{m^2r^3(9\alpha_3^2-16\alpha_4)}+2(1+\frac{1}{3\alpha_3}))^2]^{\frac{1}{2}}dr},\nonumber\\
\phi^i&=&(\frac{2}{3\alpha_3}+1)x^i,
\end{eqnarray}
where $r_s$ and $S$ are integral constants. The metric (\ref{48}) differs from the Schwarzschild-de Sitter solution by an additional term $\frac{\ln{r}}{r}$. The solution (\ref{46}) or (\ref{48}) may possess an event horizon depending upon the parameters $r_s$ and $S$, which becomes a candidate for the modified black hole. In other words, such black holes can be described by two physical parameters: the Schwarzschild radius $r_s$ and the scalar charge $S$, so that they are dubbed furry black holes. If these solutions possess an event horizon, we can show that the singularity of $I^{ab}$ is absent at the horizon (see Appendix A).

\section{Charged black hole}
The Schwarzschild metric and furry metric, derived in Sec. VI, are solutions of the vacuum gravitational field equations (\ref{5}). In this section, we derive the solution of a spherically symmetric charged body. Such a metric is a solution of the modified Einstein equations (\ref{5}) with a nonvanishing energy-momentum tensor $T^{(m)}_{\mu\nu}$ which arises from the electromagnetic field.

A spherically symmetric electromagnetic potential $A_\mu$ will have the following vanishing components: $A_2=A_3=0$. Because $A_\mu$ still has the gauge freedom, the only nonvanishing component left of the electromagnetic vector potential is $A_0$ under a suitable gauge. The nonvanishing  components of electromagnetic field tensor $F_{\mu\nu}$ are then given by
\begin{eqnarray}\label{49}
F_{01}=\frac{\partial A_0}{\partial r}=-F_{10},
\end{eqnarray}
and the energy-momentum tensor
\begin{eqnarray}\label{50}
T^{(m)\ \mu}_{\quad\quad\ \nu}=\frac{Q^2}{8\pi r^4}
\left(\begin{array}{cccc}
-1&0&0&0\\
0&-1&0&0\\
0&0&1&0\\
0&0&0&1
\end{array}\right).
\end{eqnarray}

\subsection{The Reissner-Nordstr\"{o}m solution}
In the case of $\beta=1$, the modified Einstein equations can be reduced to
\begin{eqnarray}
\frac{(b^2)'}{r}&+&\frac{b^2-1}{r^2}+m^2(-[\sqrt{\mathbf{\Sigma}_2}]+2)+\frac{Q^2}{r^4}=0,\label{51}\\
\frac{(b^2)''}{2}&+&\frac{(b^2)'}{r}+m^2([\sqrt{\mathbf{\Sigma}_2}]-2)-\frac{Q^2}{r^4}=0.\label{52}
\end{eqnarray}
From (\ref{48}) and (\ref{49}), we have
\begin{equation}\label{53}
\frac{(b^2)''}{2}+\frac{2(b^2)'}{r}+\frac{b^2-1}{r^2}=0,
\end{equation}
and
\begin{equation}\label{54}
b^2=1-\frac{r_s}{r}+\frac{c}{r^2},
\end{equation}
where $r_s$ and $c$ are integral constants. Substituting (\ref{54}) into (\ref{51}) or (\ref{52}), we obtain $c=Q^2$ and the analytical expression of $h(r)$. We finally obtain the solution
\begin{eqnarray}\label{55}
b^2&=&\frac{1}{a^2}=1-\frac{r_s}{r}+\frac{Q^2}{r^2},\nonumber\\
\phi^0&=&t\pm h(r),\\
\phi^i&=&x^i,\nonumber
\end{eqnarray}
where
\begin{eqnarray}\label{55b}
h(r)=\frac{r_s}{2}\ln{(r^2-r_sr+Q^2)}
+\left\{
    \begin{array}{ll}
    \frac{r_s^2-2Q^2}{2\sqrt{r_s^2-4Q^2}}\ln{\frac{2r-r_s-\sqrt{r_s^2-4Q^2}}{2r-r_s+\sqrt{r_s^2-4Q^2}}}, &\mbox{\quad\quad $r_s^2>4Q^2$,}\\
    \frac{r_s^2}{2(r_s-2r)}, &\mbox{for\ \  $r_s^2=4Q^2$,}\\
    \frac{r_s^2-2Q^2}{\sqrt{4Q^2-r_s^2}}\arctan{\frac{2r-r_s}{\sqrt{4Q^2-r_s^2}}},&\mbox{\quad\quad $r_s^2<4Q^2$.}
    \end{array}
    \right.
\end{eqnarray}
Equation (\ref{55}) is nothing but the Reissner-Nordstr\"{o}m solution of GR for the metric; nevertheless, it should be accompanied by nontrivial backgrounds for the St\"{u}ckelberg fields. If $r_s=Q=0$, (\ref{55}) goes back to the Minkowski  metric and the St\"{u}ckelberg fields in the unitary gauge. If and only if $r_s^2\geq4Q^2$, the solution (\ref{55}) is provided with the event horizon which passes into a candidate of the black hole. From (\ref{33}) and (\ref{55}) in the $r_s^2>4Q^2$ case, we have
\begin{eqnarray}\label{56}
&I^{00}&=-(1+\frac{r_s}{r}-\frac{Q^2}{r^2}),\nonumber\\
&I^{0i}&=I^{i0}=\pm(\frac{r_s}{r}-\frac{Q^2}{r^2})n^i,\\
&I^{ij}&=\delta^{ij}-(\frac{r_s}{r}-\frac{Q^2}{r^2})n^in^j.\nonumber
\end{eqnarray}
Therefore, the singularities in the invariant $I^{ab}$ are absent except the physical singularity $r=0$.

\subsection{New-charged solution}
The modified Einstein equations (\ref{5}) with the electromagnetic energy-momentum tensor $T^{(m)}_{\mu\nu}$ in the case of $\beta=\frac{2}{3\alpha_3}+1$ can be written as
\begin{eqnarray}
\frac{(b^2)'}{r}&+&\frac{b^2-1}{r^2}+m^2(\lambda_1[\sqrt{\mathbf{\Sigma}_2}]+\lambda_2)+\frac{Q^2}{r^4}=0,\label{57}\\
\frac{(b^2)''}{2}&+&\frac{(b^2)'}{r}+m^2(\lambda_3[\sqrt{\mathbf{\Sigma}_2}]+\lambda_4)-\frac{Q^2}{r^4}=0,\label{58}
\end{eqnarray}
where $\lambda_1, \cdots, \lambda_4$, see (\ref{45}). Combining (\ref{57}) with (\ref{58}), we have
\begin{equation}\label{59}
(b^2)''+\frac{2}{r}(1-\frac{\lambda_3}{\lambda_1})(b^2)'-\frac{2}{r^2}\frac{\lambda_3}{\lambda_1}b^2+2(\lambda_4-\frac{\lambda_2\lambda_3}{\lambda_2})+2(\frac{\lambda_3}{\lambda_1}-1)\frac{Q^2}{r^4}=0.
\end{equation}

There are three subcases which correspond to distinct types of solutions: (i) $\alpha_4\neq\frac{5}{16}\alpha_3^2-\frac{\alpha_3}{6}$ and $\alpha_4\neq\frac{3}{8}\alpha_3^2-\frac{\alpha_3}{8}$, (ii) $\alpha_4=\frac{5}{16}\alpha_3^2-\frac{\alpha_3}{6}$ and (iii) $\alpha_4=\frac{3}{8}\alpha_3^2-\frac{\alpha_3}{8}$. In the subcase (i), Eqs. (\ref{57}) and (\ref{59}) have the solution as follows
\begin{eqnarray}\label{60}
b^2&=&\frac{1}{a^2}=1-\frac{r_s}{r}-\frac{S}{r^\lambda}+\frac{4m^2r^2}{27\alpha_3^2}+\frac{\nu}{r^2},\nonumber\\
\phi^0&=&t\pm\int{[(a^2+1)^2-a^2(\frac{3\alpha_3^2}{m^2(9\alpha_3^2-16\alpha_4)}(\frac{S(1-\lambda)}{r^{\lambda+2}}+\frac{\nu-Q^2}{r^4})+2(1+\frac{1}{3\alpha_3}))^2]^{\frac{1}{2}}dr},\nonumber\\
\phi^i&=&(\frac{2}{3\alpha_3}+1)x^i,
\end{eqnarray}
where $r_s$ and $S$ are integral constants, $\lambda$ see (\ref{44}) and
\begin{equation}\label{61}
\nu=\frac{(3\alpha_3^2+2\alpha_3)Q^2}{2(3\alpha_3^2-\alpha_3-8\alpha_4)}.
\end{equation}
This metric differ from the Schwarzschild-de Sitter solution by two additional powerlike terms, $r^{-\lambda}$ and $r^{-2}$.

In the subcase (ii), the equations (\ref{57}) and (\ref{59}) have the solution as follows
\begin{eqnarray}\label{62}
b^2&=&\frac{1}{a^2}=1-\frac{r_s}{r}-\frac{S\ln{r}}{r}+\frac{4m^2r^2}{27\alpha_3^2}+\frac{3Q^2}{r^2},\nonumber\\
\phi^0&=&t\pm\int{[(a^2+1)^2-a^2(\frac{3\alpha_3^2}{m^2(9\alpha_3^2-16\alpha_4)}(\frac{S}{r^3}+\frac{2Q^2}{r^4})+2(1+\frac{1}{3\alpha_3}))^2]^{\frac{1}{2}}dr},\nonumber\\
\phi^i&=&(\frac{2}{3\alpha_3}+1)x^i.
\end{eqnarray}
The metric (\ref{62}) differs from the Schwarzschild-de Sitter solution by two additional terms, $\frac{\ln{r}}{r}$ and $r^{-2}$.

In the subcase (iii), Eqs. (\ref{57}) and (\ref{59}) have the solution as follows
\begin{eqnarray}\label{63}
b^2&=&\frac{1}{a^2}=1-\frac{r_s}{r}+\frac{Q^2}{r^2}-\frac{2Q^2\ln{r}}{r^2}+\frac{4m^2r^2}{27\alpha_3^2},\nonumber\\
\phi^0&=&t\pm\int{[(a^2+1)^2-a^2(\frac{6Q^2\alpha_3^2(1-\ln{r})}{m^2(9\alpha_3^2-16\alpha_4)r^4}+2(1+\frac{1}{3\alpha_3}))^2]^{\frac{1}{2}}dr},\nonumber\\
\phi^i&=&(\frac{2}{3\alpha_3}+1)x^i.
\end{eqnarray}
The metric (\ref{63}) is distinguished from the Reissner-Nordstr\"{o}m-de Sitter solution by an additional term $\frac{\ln{r}}{r^2}$.

In all three subcases, the singularities of $I^{ab}$ are absent except the physical singularity $r=0$. These solutions may possess event horizon so that they are candidates for the modified black hole. The event horizon of solution (\ref{60}) or (\ref{62}) depends on three physical parameters: the Schwarzschild radius $r_s$, electric charge $Q$ and scalar charge $S$, so that they are charged furry black holes. However, the black hole solution (\ref{63}) only depends on the electric charge.
\begin{table*}
\small \caption{New exact solutions and their properties in dRGT with $\alpha_3\neq-\frac{3}{2}$ and $0$.}
\begin{tabular}{cccc}
 \hline
Solution &Scalar charge &Electric charge   &Parameter $\alpha_4$\\ \hline
(\ref{46}) &Yes            &No               &$\alpha_4\neq\frac{5}{16}\alpha_3^2-\frac{\alpha_3}{6}$ \\
(\ref{48}) &Yes            &No               &$\alpha_4=\frac{5}{16}\alpha_3^2-\frac{\alpha_3}{6}$ \\
(\ref{60}) &Yes            &Yes              &$\alpha_4\neq\frac{5}{16}\alpha_3^2-\frac{\alpha_3}{6}$ and $\alpha_4\neq\frac{3}{8}\alpha_3^2-\frac{\alpha_3}{8}$ \\
(\ref{62}) &Yes            &Yes              &$\alpha_4=\frac{5}{16}\alpha_3^2-\frac{\alpha_3}{6}$\\
(\ref{63}) &No            &Yes               &$\alpha_4=\frac{3}{8}\alpha_3^2-\frac{\alpha_3}{8}$\\
\hline
\end{tabular}\\
\end{table*}

\section{Conclusion and discussion}
In GR, the spherically symmetric solution to the Einstein equation is a benchmark, and its massive deformation also plays a crucial role in dRGT. In this work, we have developed a study of the spherically symmetric solutions in dRGT if the St\"{u}ckelberg fields are taken as a hedgehog configuration $\phi^0=t+h(r)$ and $\phi^i=\phi(r)x^i/r$. Under the hedgehog configuration and the static spherically symmetric metric (\ref{10}), the self-consistency imposes $\phi(r)=\beta r$: $\beta=1$ or $\beta=\frac{2}{3\alpha_3}+1$. Note that there is only $\beta=1$ from (\ref{30}) in the case of $\alpha=-3/2$ or $0$.

On the premise of $\beta=1$, we showed that there is always the Schwarzschild solution and Reissner-Nordstr\"{o}m solution in dRGT with two free parameters. Furthermore, we can prove that the singularities in the invariant $I^{ab}$ are absent except the physical singularity $r=0$, so that these solutions of massive gravity may be regarded as candidates for the black hole in dRGT. That is, we are able to reproduce the behavior of GR in the static spherically symmetric case without constraint for the parameters $\alpha_3$ and $\alpha_4$. It would be interesting to consider the connection between dRGT and GR based on this result. In particular we consider a point source of mass $M$ localized at $r=0$. In the Newtonian approximation of GR, the gravitational potential mediated by the point source is $\Psi(r)=-\frac{r_s}{r}$ where $r_s$ is the Schwarzschild radius associated with the source. The helicity-0 mode of the graviton also contributes to the gravitational potential with an additional amount $\delta\Psi$ in dRGT. However, the nontrivial configuration of the St\"{u}ckelberg fields also contributes an additional term which is exactly canceled by $\delta\Psi$. Within such context, the Newtonian potential is still $\Psi(r)$ in dRGT.

We also presented five exact solutions in dRGT for the case of $\beta=\frac{2}{3\alpha_3}+1$ and $\alpha_3\neq-\frac{3}{2}$ and $0$ (see Table I). These solutions have also shown that the singularities in the invariant $I^{ab}$ are absent except the physical singularity $r=0$, so that they may be regarded as candidates of black hole in dRGT. Such black holes [except (\ref{63})] contain an additional physical parameter (scalar charge $S$); therefore, they are dubbed furry black holes.

In addition, one may be anxious that the scalar perturbations on these backgrounds are infinitely strongly coupled in light of the results of Ref.\cite{Rham4}. It was found that the de Sitter background has infinitely strongly coupled fluctuations in the decoupling limit for the parameters chosen as $9\alpha_3^2+3\alpha_3-12\alpha_4+1=0$ \cite{Rham4}. In this work, we have $\pi^0=-h(r)$ and $\pi^i=(1-\beta)x^i$. From (\ref{9}), we obtain the vector mode $A^0=-\frac{\Lambda^3}{m}h(r)$ and $A^i=0$ which is different from that studies in \cite{Rham4}.

Finally, we can also discuss the cosmic acceleration using our method in this work and will do so in a forthcoming paper.

\begin{acknowledgments}
This work is supported by National Science Foundation of China Grant. No. 11205102 and the Innovation Program of Shanghai Municipal Education Commission (12YZ089).
\end{acknowledgments}

\appendix

\section{NEW BASIC INVARIANT $I^{ab}$}
The existence of St\"{u}ckelberg fields leads to a basic invariant $I^{ab}=g^{\mu\nu}\partial_\mu\phi^a\partial_\nu\phi^b$ in dRGT. Under ansatz
 (\ref{10}), $I^{ab}$ can be written as
 \begin{eqnarray}\label{A1}
&I^{00}&=\frac{h'^2}{a^2}-\frac{1}{b^2},\nonumber\\
&I^{0i}&=I^{i0}=\frac{\phi'h'}{a^2}n^i,\\
&I^{ij}&=\frac{\phi^2}{r^2}\delta^{ij}+(\frac{\phi'^2}{a^2}-\frac{\phi^2}{r^2})n^in^j,\nonumber
\end{eqnarray}
where $(n^1,n^2,n^3)=(\sin{\theta}\cos{\varphi},\sin{\theta}\sin{\varphi},\cos{\theta})$. In the case of $\phi=\beta r$, we have the Schwarzschild metric; if it were the solution of dRGT in the unitary gauge $\phi^a=x^\mu\delta^a_\mu$ would be singular at $r=r_s$ according to (\ref{A1}):
\begin{eqnarray}\label{A2}
&I^{00}&=\frac{1}{1-\frac{r_s}{r}},\nonumber\\
&I^{0i}&=I^{i0}=0,\\
&I^{ij}&=\delta^{ij}.\nonumber
\end{eqnarray}

For the Schwarzschild metric in the nonunitary gauge (\ref{40}), we have
\begin{eqnarray}\label{A3}
&I^{00}&=-(1+\frac{r_s}{r}),\nonumber\\
&I^{0i}&=I^{i0}=\pm\frac{r_s}{r}n^i,\\
&I^{ij}&=\delta^{ij}-\frac{r_s}{r}n^in^j,\nonumber
\end{eqnarray}
so the coordinate singularity at $r=r_s$ is absent. Note that there is still the physical singularity $r=0$ in the invariant $I^{ab}$ just as the ones usually encountered in GR (Ricci scalar, Ricci tensor square, Riemann tensor square, etc.).

Generally, we can prove that the singularities in the invariant $I^{ab}$ are absent except the physical singularity $r=0$ under the self-consistent ansatz (\ref{32}). In fact, from the modified Einstein equations in empty space we have
\begin{equation}\label{A4}
[\sqrt{\mathbf{\Sigma}_2}]^2=\frac{1}{\lambda_1^2r^4}[(b^2)'+(b^2-1)+\lambda_2r^2]^2.
\end{equation}
On the other hand, we have
\begin{equation}\label{A5}
h'^2=\frac{(1+\beta b^2)^2-b^2[\sqrt{\mathbf{\Sigma}_2}]^2}{b^4}.
\end{equation}
To combine (\ref{A1}), (\ref{A4}) and (\ref{A5}), we obtain $I^{ab}$ as follows
\begin{eqnarray}\label{A6}
&I^{00}&=2\beta+\beta^2b^2-[\sqrt{\mathbf{\Sigma}_2}]^2,\nonumber\\
&I^{0i}&=I^{i0}=\pm\beta\sqrt{(1+\beta b^2)^2-b^2[\sqrt{\mathbf{\Sigma}_2}]^2}n^i,\\
&I^{ij}&=\beta^2\delta^{ij}+\beta^2(b^2-1)n^in^j,\nonumber
\end{eqnarray}
So the singularities are absent except the physical singularity $r=0$, since coordinate singularity appears only in the negative power term of $b^2$.

\end{document}